\documentclass[a4paper]{article}

\usepackage{INTERSPEECH2021}
\usepackage{color}
\usepackage{url}
\usepackage{multirow,array}

\usepackage{bm}
\usepackage{amssymb,amsmath,graphicx,bbm,color,booktabs}
\usepackage{amsopn}

\newcommand{\ve}{\bm{e}}

\newcommand{\vp}{\bm{p}}       \newcommand{\vph}{\hat{\bm{p}}}

\newcommand{\vv}{\bm{v}}               
               
\newcommand{\vx}{\bm{x}}       \newcommand{\vxh}{\hat{\bm{x}}}        
               
\newcommand{\vz}{\bm{z}}               

    \newcommand{\Xc}{\mathcal{X}}

\newcommand{\R}{\mathbb{R}}

\renewcommand{\eqref}[1]{Eq.~(\ref{#1})}

\title{Speech Resynthesis from Discrete \\ Disentangled Self-Supervised Representations}
\name{
Adam Polyak$^{1,2}$\sthanks{ \hspace{0.1cm} The contribution of Adam Polyak is part of a Ph.D. thesis
research conducted at Tel Aviv University.}, Yossi Adi$^{1}$, Jade Copet$^{1}$, Eugene Kharitonov$^{1}$, Kushal Lakhotia$^{1}$, Wei-Ning Hsu$^{1}$, Abdelrahman Mohamed$^{1}$, Emmanuel Dupoux$^{1,3}$}
\address{
  $^1$Facebook AI Research, USA \\
  $^2$Tel-Aviv University, Israel \\  $^3$EHESS, France}
\email{adampolyak@fb.com}

\begin{document}

\maketitle

\begin{abstract}
\end{abstract}
We propose using self-supervised discrete representations for the task of speech resynthesis. To generate disentangled representation, we separately extract low-bitrate representations for speech content, prosodic information, and speaker identity. This allows to synthesize speech in a controllable manner. We analyze various state-of-the-art, self-supervised representation learning methods and shed light on the advantages of each method while considering reconstruction quality and disentanglement properties. Specifically, we evaluate the F0 reconstruction, speaker identification performance (for both resynthesis and voice conversion), recordings' intelligibility, and overall quality using subjective human evaluation. Lastly, we demonstrate how these representations can be used for an ultra-lightweight speech codec. Using the obtained representations, we can get to a rate of 365 bits per second while providing better speech quality than the baseline methods. Audio samples can be found under the following link: {\color{magenta} \url{speechbot.github.io/resynthesis}}.


% our contribution: 
% 1) Demonstrate the usage of SSL units for synthesis purposes (no Tacotron model)
% 2) Directly evaluate SSL units from a TTS point of view, Voice Conversion, and F0 manipulation
% 3) Demonstrate the usage of such units for speech codec

\vspace{-0.1cm}
\section{Introduction}
\vspace{-0.1cm}
Learning unsupervised speech representations, both continuous and discrete, has seen a significant leap in performance following the recent success of Self-Supervised Learning (SSL) methods~\cite{oord2018representation, schneider2019wav2vec, baevski2020wav2vec, hsu2020hubert}. In the self-supervised setting, unlabeled inputs define an auxiliary task that can generate pseudo-labeled training data. This data can then be used to train a model using supervised techniques. %The learned representations can then be used for downstream tasks with or minimal amount of supervised data. For example, with 10 minutes of transcribed speech and 53K hours of untranscribed speech, wav2vec 2.0 enables speech recognition models at a word error rate (WER) of 8.6 percent on noisy speech and 5.2 percent on clean speech on the standard LibriSpeech benchmark~\cite{baevski2020wav2vec}.
The learned representations are often used for downstream tasks with a minimal amount of supervised data. For example, learning  speech recognition with only 10 minutes of transcribed speech and 53K hours of untranscribed speech as in wav2vec 2.0~\cite{baevski2020wav2vec} and HuBERT~\cite{hsu2020hubert}. The learned self-supervised discrete representations also showed impressive performance on the conditional and unconditional spoken generative language modeling (GSLM) task~\cite{lakhotia2021generative}.

Despite its success, most studies on SSL for speech are focused on generating and evaluating the quality of the learned representations in the context of Automatic Speech Recognition (ASR). It remains unclear how suitable these representations are for speech synthesis. Moreover, in the context of expressive and controllable generation, it is unknown to what extent the speaker identity and F0 information are encoded in the learned representations. 

\begin{figure}[t!]
\centering
\includegraphics[scale=0.4,trim={50 20 70 20}]{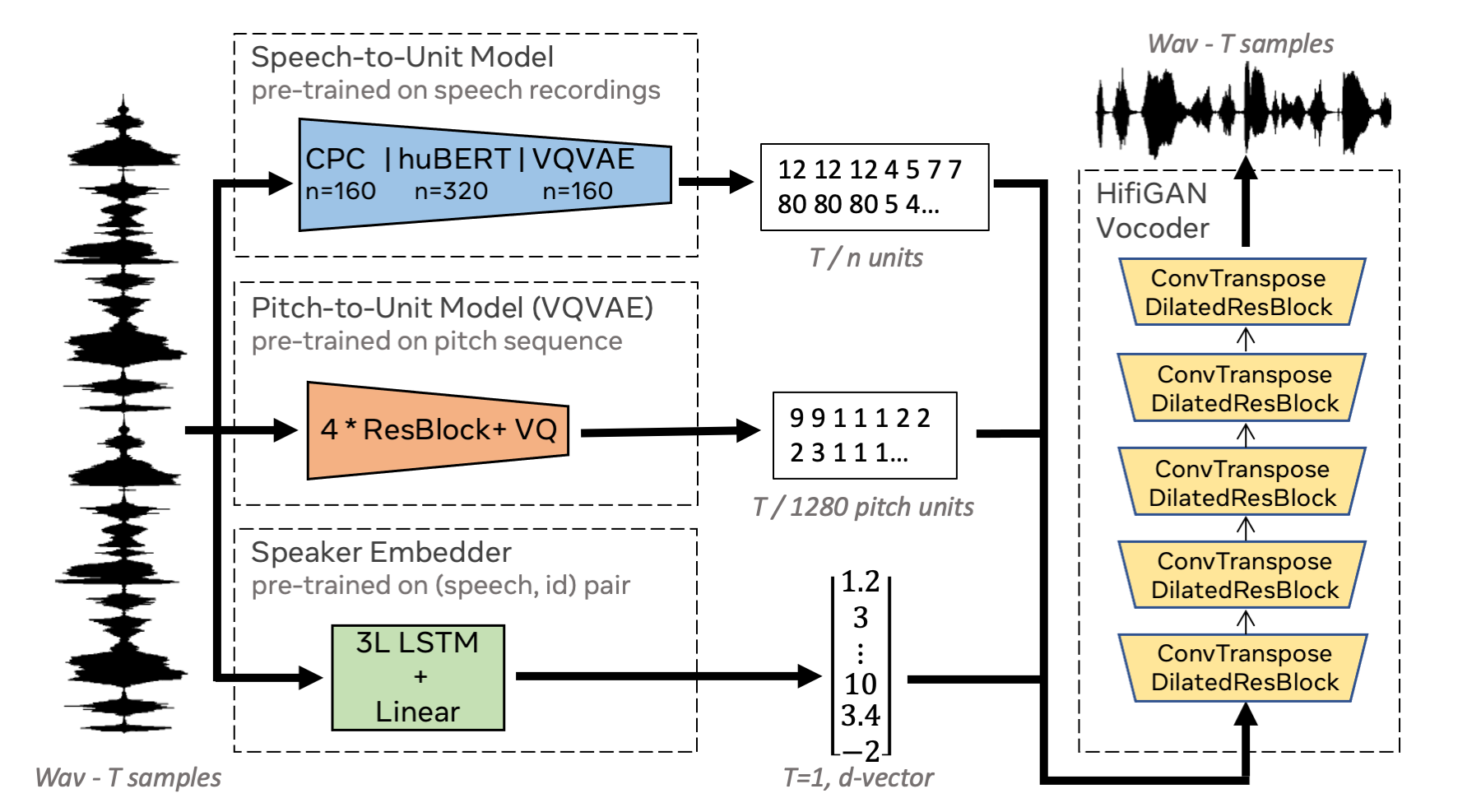}
\caption{\textbf{The overall proposed speech resynthesis architecture.} Three parallel encoders extract discrete representations from the raw input signal. These are then being used as a conditioning to reconstruct the signal using a decoder network.} 
\label{fig:arch}
\vspace{-0.6cm}
\end{figure}

% Traditionally, speech synthesis and text-to-speech models get as input textual features and generate Mel-spectrogram autoregressively. 
Traditionally, speech synthesis and text-to-speech models produce Mel-spectrogram autoregressively given textual features as input.
Next, a vocoder is applied to reconstruct the phase from the Mel-spectrogram (e.g., Griffin-Lim~\cite{griffin1984signal}, WaveNet~\cite{oord2016wavenet}, WaveGlow~\cite{waveglow}, or HiFi-GAN~\cite{kong2020hifi}). In this study, we suggest using the learned speech units as an input to a vocoder module with no spectrogram estimation. We additionally augment the learned units with quantized F0 representation and a global speaker embedding. Figure~\ref{fig:arch} presents the overall proposed method.
% This allows to evaluate the encoded properties in the learned units when considering speech content, speaker identity, and F0 information, as well as better control the audio synthesis. 
This allows the evaluation of the learned units with respect to speech content, speaker identity, and F0 information, as well as better control the audio synthesis. 
We experiment with signal reconstruction, voice conversion, and F0 manipulation using several datasets and encoder models. Finally, equipped with our previous findings, we demonstrate how the learned units can function as an ultra-lightweight speech codec. Following the proposed method, we can reach an encoding rate of 365 bits per second~(bps) while being superior to the baseline methods by a significant margin, including lightweight and heavyweight codecs. 

% \paragraph*{Our Contribution:}
\noindent{\bf Our Contribution:\quad}
% for the first time 
(i) We demonstrate the usage of discrete speech units, learned in a self-supervised manner, for high-quality synthesis purposes (no Mel-spectrogram estimation); (ii) We provide an extensive evaluation of the SSL speech units from a synthesis point of view, i.e., signal reconstruction, voice conversion, and F0 manipulation; (iii) We build an ultra-lightweight speech codec from the obtained speech units.

\vspace{-0.1cm}
\section{Related Work}
\vspace{-0.1cm}
% start with related work regarding speech representations (VAE, CPC, WAV2VEC, HuBERT, etc.) and the tacl paper, and the zero speech challenge 
% \paragraph*{Unsupervised Speech Representation Learning.} 
\noindent{\bf Unsupervised Speech Representation Learning\quad}
Studies on unsupervised speech representation learning can roughly be divided into reconstruction and self-supervised learning methods. 
%  The common approach for signal reconstruction is auto-encoding, 
Auto-encoding is the common approach for signal reconstruction,
where speech is first encoded into a low-dimensional latent representation, and then decoded back to speech. Various constraints can be imposed on the encoded space, such as temporal smoothness~\cite{ebbers2017hidden}, discreteness~\cite{van2017neural}, and hierarchy~\cite{hsu2017unsupervised}.

SSL methods have shown remarkable results for ASR~\cite{schneider2019wav2vec, baevski2020wav2vec}, phoneme segmentation~\cite{kreuk2020self}, and GSLM~\cite{lakhotia2021generative}. The authors in~\cite{oord2018representation, schneider2019wav2vec} suggested training a convolutional neural network to distinguish true future samples from random distractor samples using a Contrastive Predictive Coding (CPC) loss function. Similar to CPC, the authors in~\cite{baevski2020wav2vec} use an encoder and a predictor, which is trained contrastively to distinguish positive and negative samples. However, unlike~\cite{schneider2019wav2vec} it discretizes and masks segments of the encoder's output. In HuBERT~\cite{hsu2020hubert}, the model is trained with a masked prediction task similar to BERT~\cite{devlin-etal-2019-bert} but with masked continuous audio signals. 

% text to speech
% describe some disentangled speech representation work 
% Controllable neural text-to-speech synthesis using intuitive prosodic features
% \paragraph*{Speech Resynthesis.}
\noindent{\bf Speech Resynthesis \quad}
Recent advancements in neural-based vocoders enabled generating natural-sounding speech and music~\cite{oord2016wavenet, prenger2019waveglow, kong2020hifi}. These are often conditioned on the log Mel-spectrogram for the generation process. 
% Learning low bitrate speech representations in an unsupervised manner is widely explored recently, mainly under the Zero-Resource Challenge~\cite{dunbar2019zero, dunbar2020zero}.
Recently, unsupervised learning of low bitrate speech representations was explored under the Zero-Resource Challenge~\cite{dunbar2019zero, dunbar2020zero}.
Vector-Quantized Variational Auto-Encoder (VQ-VAE)~\cite{van2017neural} employs a learned fixed-sized codebook, decoded by a WaveNet model for speech synthesis. In~\cite{eloff2019unsupervised} the authors proposed a VQ-VAE model followed by an FFTNet vocoder model~\cite{jin2018fftnet}. The authors in~\cite{tjandra2020transformer} suggested using transformer~\cite{trans} together with a VQ-VAE model for unsupervised unit discovery, and in~\cite{van2020vector} they combine vector quantization with contrastive predictive coding for acoustic unit discovery. The study by~\cite{zhao2020improved} is the closest to our work. In which, the authors suggest training a VQ-VAE with two encoders, one for the waveform and the other for F0. The authors demonstrated how such modeling improves overall generation quality. In contrast, we study SSL-based speech encoders and empirically show these representations are better disentangled, and apply them as an ultra-low bitrate speech codec. Another line of work suggests using intermediate representations obtained from an ASR acoustic model. These representations are being used together with the identity and prosodic information for voice conversion~\cite{polyak2019tts, polyak2020unsupervised, polyak2021high}. Unlike all of the above, we suggest synthesizing speech directly from the discrete units. Moreover, the resynthesis process sheds light on the encoded information in each of the evaluated representations. 

% \paragraph*{Speech Codec.}
% \smallskip
\noindent{\bf Speech Codec \quad}
% lastly, some speech codec work (both AI and non-AI based)
Speech codecs typically employ a carefully hand-engineered pipeline combining an encoder and a decoder 
% that considers the physics of speech production 
which is influenced by speech production physics
to remove redundancies in the data and yield a compact bitstream. Low bitrate parametric speech codecs have long been studied~\cite{atal1971speech}, but their quality has been severely limited. Despite some advances~\cite{griffin1985new, mccree19962}, modeling the excitation signal has remained a challenge. Neural speech codecs have been recently proposed and demonstrated promising results~\cite{kleijn2018wavenet, lim2020robust}. 
In~\cite{valin2019real} an LPCNet~\cite{valin2019lpcnet} vocoder was conditioned on hand-crafted features and a uniform quantizer. In~\cite{garbacea2019low} a WaveNet model was conditioned on discrete units obtained from a VQ-VAE model, while in~\cite{skoglund2019improving} the Opus codec~\cite{valin2012definition} was fed to WaveNet. % module.
\vspace{-0.1cm}
\section{Method}
\vspace{-0.1cm}
\label{sec:method}
The proposed architecture is comprised of three pre-trained and fixed encoders, namely: (i) content encoder; (ii) F0 encoder; (iii) speaker identity encoder; and a decoder network. The first two encoders extract discrete representation from the raw audio while the latter extracts a single global representation. The overall architecture is depicted in Figure~\ref{fig:arch}. 

\vspace{-0.2cm}
\subsection{Encoders}
\vspace{-0.1cm}
%In the following subsections we provide a detailed description for each of the modules.  
% \paragraph*{Content Encoder.}
Denote the domain of audio samples by $\Xc \subset \R$. The representation for a raw signal is therefore a sequence of samples $\vx = (x_1,\ldots, x_T)$, where  $x_t\in\Xc$ for all $1\leq t \leq T$. 

\noindent{\bf Content Encoder \quad} 
% Consider an encoder network, $E_c$, that gets as input the speech utterance and outputs 
The input to a content encoder network, $E_c$, is a speech utterance, $\vx$, and the output is 
a sequence of spectral representations sampled at a low frequency as follows $E_c(\vx) = (\vv_1, \dots, \vv_{T'})$. We evaluated three state-of-the-art unsupervised representation learning functions as $E_c$. Specifically, we experimented with: (i) CPC~\cite{oord2018representation} which attempts to predict the future states of the encoder based on the past and optimizes a contrastive loss comparing the actual future from that of random sequences; (ii) HuBERT~\cite{hsu2020hubert} which was trained with a masked prediction task similar to BERT~\cite{devlin-etal-2019-bert} on masked continuous audio signals as inputs. The targets are obtained through clustering of raw speech features or learned features from earlier iterations; (iii) and VQ-VAE~\cite{van2017neural} which performs similarly to a Variational Auto Encoder~\cite{kingma2013auto} where the encoder's output is discrete rather than continuous. 

Since the representations learned by CPC and HuBERT are continuous, a k-means algorithm is applied over the models' outputs to generate discrete units, denoted as $\vz_c = (z_1,\ldots,z_{L})$. Each element $z_i$ in $\vz_c$ is a positive integer, $z_i\in\{0,1,..,K\}$ for $1\le i \le L$, where $K$ is the number of discrete units. We did not follow the same approach for VQ-VAE as its representations are already quantized.

\noindent{\bf F0 Encoder \quad} 
To generate low frequency discrete F0 representation, $\vz_{F_0}=(z_1, \dots, z_{L'})$, a separate encoder, $E_{F_0}$, is applied over the F0 extracted from the input signal. Each element in $\vz_{F_0}$ is an integer $z_s\in\{0,1,..,K'\}$, where $K'$ is the encoder dictionary size. The YAAPT~\cite{yaapt} algorithm is used to extract the F0 from the input signal, $\vx$, generating~$\vp = (p_1, \dots, p_{T'})$. 

$E_{F_0}$ is trained using the VQ-VAE framework. VQ-VAE employs a convolutional encoder, $E_{F_0}$,  a bottleneck with a learned codebook $C = (\ve_1, \dots, \ve_{K'})$, where each item in $C$ is a 128-dimensional vector, and a decoder $D_{F_0}$. The encoder extracts a sequence of latent vectors $E_{F_0}(\vp)=({\bf{h}}_1, \dots, {\bf{h}}_{L'})$ 
from the raw audio, where ${\bf{h}}_i \in \R^{128}$, for all $1 \le i \le L'$. 
Then, the bottleneck maps each latent vector to its nearest vector in the codebook $C$. The embedded latent vectors are then being fed into the decoder $D_{F_0}(\ve_{z_1}, \dots, \ve_{z_{L'}}) = \vph$ which reconstructs the original F0 signal. Similar to~\cite{jukebox}, we use Exponential Moving Average updates to learn the codebook and employ random restarts for unused embeddings. 

% Overall, the VQ-VAE is trained by minimizing the following loss:
% \begin{equation}
% \begin{aligned}
%     &L(E_{F_0}, C, D_{F_0}) = L_{recon} + \beta L_{commit}, \\
%     &L_{recon}(E_{F_0}, C, D_{F_0}) = \frac{1}{T'}\sum_{t=1}^{T'} \| p_t - D_{F_0}(e_{t}) \|^2_2 \\
%     &L_{commit}(E_{F_0}, C) = \frac{1}{L'}\sum_{s=1}^{L'} \| h_s - \text{sg}[e_{z_s}] \|^2_2,
% \end{aligned}
% \end{equation}
% where $\texttt{sg}$ is the stop gradient operation, detaching the embedding from the loss backpropagation. %The discrete representation is achieved by taking the indices of the embedded latent vectors $\ve$.

% The reconstruction term reduces the distance between the input and the generated signal, while the commitment term ensures the latent representation remains close to the codebook vectors. 
To generate $\vz_{F0}$, we use the indices of the mapped latent vectors rather than the vectors. 

% \paragraph*{Speaker Encoder.}
% \smallskip
\noindent{\bf Speaker Encoder \quad} 
\label{sec:spkr}
Lastly, a speaker encoder, $E_{spk}$, is used to extract speaker embedding. A pre-trained speaker verification model similar to the one proposed in~\cite{heigold2016end} is used. Formally, $E_{spk}$ gets as input the speech utterance, $\vx$, extracts the Mel-spectrogram and outputs a \emph{d-vector} speaker representation, denoted as $\vz_{spk} \in \R^{256}$. We experimented with learning the speaker embedding via a lookup-table. Although such method performs slightly better, it is limited to speakers seen during training.

\begin{table}[t!]
\centering
\caption{Speech resynthesis results. MOS results are reported with 95\% confidence interval. Except MOS for all other evaluation metrics the lower is the better. We did not report EER for LJ since it is a single speaker dataset.}
\label{tab:recon}
\resizebox{\columnwidth}{!}{
\begin{tabular}{c@{~} | c@{~} | c@{~} c@{~} | c@{~}c@{~} | c@{~} | c@{~}}
\toprule
\multirow{2}{*}{Dataset} & \multirow{2}{*}{Method} & \multicolumn{2}{c|}{Content} & \multicolumn{2}{c|}{F0} &  Speaker & Overall Quality \\
\cmidrule{3-8}
\cmidrule{3-8}
 &  & PER~$\downarrow$ & WER~$\downarrow$ & VDE~$\downarrow$ & FFE~$\downarrow$ &  EER~$\downarrow$ & MOS~$\uparrow$ \\
\midrule
\multirow{4}{*}{LJ}
& GT        & 6.93 & 5.60 & -- & -- & -- & 4.33$\pm$0.20 \\
\cmidrule{2-8}
% & {\color{red}ASR-TTS}   & {\color{red}9.78}  & {\color{red}7.43} & -- & -- & -- &  \\
& CPC       & 9.66  	& 8.51 		& 13.48 		& 15.19 		& -- & 3.31$\pm$0.33 \\
& HuBERT    & \bf 9.52  & \bf 6.96 	& 13.09 		& 15.00 		& -- & \bf 3.66$\pm$0.33\\
& VQ-VAE    & 12.77 	& 8.85 		& \bf 7.19 	& \bf 8.54 	& -- & \bf 3.66$\pm$0.31\\
\midrule 
\multirow{4}{*}{VCTK} 
& GT        & 17.16 & 4.32 & -- & -- & 3.25 & 4.08$\pm$0.66\\
\cmidrule{2-8}
% & {\color{red}ASR-TTS}   & {\color{red}29.75} & {\color{red}8.93} & -- & -- & {\color{red}7.27} &  \\
& CPC       & 23.01 	& 14.49 		& 10.56 		& 11.13 		& \bf 4.25 	& 3.33$\pm$0.61\\
& HuBERT    & \bf 19.66 & \bf 11.44 & 9.77  		& 10.43 		& 5.79 		& \bf 3.41$\pm$0.66\\
& VQ-VAE    & 31.97 	& 19.80 		& \bf 5.20 	& \bf 5.59 	&  4.28 		& 3.39$\pm$0.58\\
\bottomrule
\end{tabular}
}
\vspace{-0.4cm}
\end{table}

\vspace{-0.2cm}
\subsection{Decoder}
\vspace{-0.1cm}
% To decode the speech signal from its encoded representation, a neural vocoder is employed. 
A neural vocoder is employed to decode the speech signal from the discrete representation.
This study considers the decoder to be a modified version of the HiFi-GAN~\cite{kong2020hifi} neural vocoder. 

The HiFi-GAN architecture is comprised of a generator, $G$, and a set of discriminators, $D$. The generator is built from a set of look-up tables~(LUT) that embed the discrete representation and a series of blocks composed of transposed convolution and a residual block with dilated layers. The transposed convolutions upsample the encoded representation to match the input sample rate, while the dilated layers increase the 
% model's 
receptive field. 

As an input, the generator receives the encoded representation $(\vz_c, \vz_{F_0}, \vz_{spk})$. The discrete content sequence, $\vz_c$, and the discrete pitch sequence, $\vz_{F_0}$, are converted to a continuous representation via $LUT_{c}$ and $LUT_{F_0}$ accordingly. The sequences are up-sampled and concatenated together. The speaker embedding, $\vz_{spk}$, is concatenated to each frame in the up-sampled sequence.  
% The concatenation is then fed to the series of transposed convolutions.

The discriminator comprises two networks, a Multi-Period Discriminator~(MPD) and a Multi-Scale Discriminator~(MSD). The MPD consists of multiple sub-discriminators operating on equally spaced samples from the input signal. The period sub-discriminators differ from each other based on the space between the samples. Similar to~\cite{kong2020hifi}, the MPD employs a total of five-period discriminators with a period hops of $[2, 3, 5, 7, 11]$. Multi-scale discriminator (MSD) employs multiple sub-discriminators operating at different scales of the input signal. Specifically, we use three scales: the original input scale, $\times2$ downsampled scale, and $\times4$ downsampled scale. Overall each sub-discriminator $D_j$ is tasked with minimizing the following,
\begin{equation}
\label{eq:discriminator}
\begin{aligned}
    &L_{adv}(D_j, G) = \sum_{\vx}|| 1 - D_j(\vxh) ||_2^2, \\
    &L_D(D_j, G) = \sum_{\vx}{[|| 1 - D_j(\vx) ||_2^2 + || D_j(\vxh) ||_2^2]},
\end{aligned}
\end{equation}
where $\vxh=G(LUT_{c}(\vz_c), LUT_{F_0}(\vz_{F_0}), \vz_{spk})$, is the resynthesized signal from the encoded representation.

% In parallel, $G$ optimizes two more auxiliary loss functions. 
Additionally, two terms are added to the loss function. 
The first one is a reconstruction term computed between the Mel-spectrogram of the input signal and the generated signal, 
% $L_{recon}(G) = \sum_{\vx}|| \phi(\vx) - \phi(\vxh) ||_1$,
\begin{equation}
\label{eq:recon}
 L_{recon}(G) = \sum_{\vx}|| \phi(\vx) - \phi(\vxh) ||_1,
\end{equation}
where $\phi$ is a spectral operator computing Mel-spectrogram. The second term is a feature-matching loss~\cite{larsen2016autoencoding} which measures the distance between discriminator activations of the real signal and those of the resynthesized signal,
% $L_{fm}(D_j, G) = \sum_{\vx}\sum_{i=1}^{R}\frac{1}{M_i}|| \psi_i(\vx) - \psi_i(\vxh) ||_1$,
\begin{equation}
\label{eq:fm}
L_{fm}(D_j, G) = \sum_{\vx}\sum_{i=1}^{R}\frac{1}{M_i}|| \psi_i(\vx) - \psi_i(\vxh) ||_1,
\end{equation}
where $\psi_i$ is an operator which extracts the activations of the discriminator $i$-th layer, $M_i$ is the number of features in layer $i$, and $R$ is the total number of layers in $D_j$. 

% The objective function of the neural vocoder, $G$, is to minimize the following,
% \begin{equation}
% \label{eq:generator}
% \begin{aligned}
%     L_{G}(D_j, G) = &L_{adv}(D_j, G) + \\ 
%     &\lambda_{fm}L_{fm}(D_j, G) + \lambda_{r}L_{recon}(G),
% \end{aligned}
% \end{equation}
% where we set $\lambda_{fm}=2$ and $\lambda_{r}=45$. 

% Finally, we update Eq.~\ref{eq:discriminator} and Eq.~\ref{eq:generator} with respect to the sub-discriminators composing discriminator, $D$:
% \begin{equation}
% \begin{aligned}
% L_G^{multi}(D, G) = &\sum_{j=1}^{J}[L_{adv}(G, D_j) + \lambda_{fm}L_{fm}(G, D_j)], \\ 
%                     & + \lambda_{r}L_{recon}(G), \\
% L_D^{multi}(D, G) = &  \sum_{j=1}^{J} L_D(G, D_j).
% \end{aligned}
% \end{equation}
% \vspace{-0.3cm}
% where $D_j$ is the $j$-th sub-discriminator of $D$.

The final loss with respect to the sub-discriminators composing discriminator $D$ and generator $G$ is:
\begin{equation}
\begin{aligned}
L_G^{multi}(D, G) = &\sum_{j=1}^{J}[L_{adv}(G, D_j) + \lambda_{fm}L_{fm}(G, D_j)], \\ 
                    & + \lambda_{r}L_{recon}(G), \\
L_D^{multi}(D, G) = &  \sum_{j=1}^{J} L_D(G, D_j),
\end{aligned}
\vspace{-0.2cm}
\end{equation}
\vspace{-0.1cm}
where we set $\lambda_{fm}=2$ and $\lambda_{r}=45$.
% where $D_j$ is the $j$-th sub-discriminator of $D$.

\begin{table}[t!]
\centering
\caption{Voice conversion \& F0 manipulation results. MOS results are reported with 95\% confidence interval. VDE, and FFE are reported for F0 manipulation while PER, WER, EER, and MOS are reported for voice conversion. Notice, for VDE, and FFE higher is the better since F0 was flattened.}
\label{tab:conv}

\resizebox{1\columnwidth}{!}{
\begin{tabular}{c@{~} | c@{~} | c@{~}c@{~} | c@{~} | c@{~} ||  c@{~}c@{~} }
\toprule
\multirow{2}{*}{Dataset} & \multirow{2}{*}{Method} & \multicolumn{4}{c||}{Voice Conversion} & \multicolumn{2}{c}{F0 Manipulation} \\
\cmidrule{3-8}
& & PER~$\downarrow$ & WER~$\downarrow$ & EER~$\downarrow$ & MOS~$\uparrow$ & VDE~$\uparrow$ & FFE~$\uparrow$ \\
\midrule
VCTK & GT  & 17.16 & 4.32 & 3.25 & 4.11$\pm$0.29 & -- & -- \\
\midrule 
\multirow{3}{*}{LJ}
% & ASR-TTS   & 50.74  & --     & 66.08 & 32.96 & 1.46 \\
& CPC       & 22.22 	& 16.11 		& 0.46 		& 3.57$\pm$0.15 		& \bf 46.68 & \bf 48.71\\
& HuBERT    & \bf 19.09 & \bf 12.23 & \bf 0.31  & \bf 3.71$\pm$0.24 & 39.20 		& 48.42\\
& VQ-VAE    & 40.88 	& 36.96 		& 9.65 		& 2.90$\pm$0.17 		& 10.54 	& 12.08 \\
\midrule 
\multirow{3}{*}{VCTK} 
% & ASR-TTS   & 68.88  & --    & 41.77 & 13.55 & 6.48 \\
& CPC       &  23.58 		& 15.98 		& \bf 4.83  &  3.42 $\pm$ 0.24 		& \bf 25.29 & \bf 26.97 \\
& HuBERT    &  \bf 20.85 	& \bf 12.72 & 6.01  		& \bf  3.58 $\pm$ 0.28 	& 23.46 	& 26.67 \\
& VQ-VAE    & 36.88  		& 29.44 		& 11.56 		& 3.08 $\pm$ 0.34 		& 7.03  	& 7.80  \\
\bottomrule
\end{tabular}}
\vspace{-0.4cm}
\end{table}

\vspace{-0.1cm}
\section{Results}
\vspace{-0.1cm}
Our results cover
% We report results for 
three different settings: (i) speech reconstruction experiments; (ii) speaker conversion and F0 manipulation; (iii) bitrate analysis with subjective tests for speech codec evaluation. We employ two datasets: LJ~\cite{ljspeech17} single speaker dataset and VCTK~\cite{vctk} multi-speaker dataset. All datasets were resampled to a 16kHz sample rate.

% \paragraph*{Implementation Details.}
% \smallskip
\noindent{\bf Implementation Details\quad} 
\label{sec:impl}
We follow the same setup as in~\cite{lakhotia2021generative}. For CPC, we used the model from~\cite{Riviere2020}, which was trained on a ``clean'' 6k hour sub-sample of the LibriLight dataset~\cite{Kahn2020,Riviere2020}. We extract a downsampled representation from an intermediate layer with a 256-dimensional embedding and a hop size of 160 audio samples. For HuBERT we used a \textsc{Base} 12 transformer-layer model trained for two iterations~\cite{hsu2020hubert} on 960 hours of LibriSpeech corpus~\cite{Panayotov2015}. 
% This model encodes every 320 raw audio samples into a 768-dimensional vector. 
This model downsamples the raw audio $\times320$ into a sequence of 768-dimensional vectors. Similarly to~\cite{lakhotia2021generative}, activations were extracted from the sixth layer.

%CPC: We use a dictionary of 100 units, leading to a bitrate of 700bps.
%HuBERT: A dictionary of 100 units is used, leading to a bitrate of 350bps. 
%VQVE: The VQ-VAE discrete code operates at a bitrate of 800bps.
% For both CPC and HuBERT, the k-means algorithm is applied to convert continuous frames to discrete codes, using the LibriSpeech clean-100h~\cite{Panayotov2015} dataset. 
For CPC and HuBERT, the k-means algorithm is trained on LibriSpeech clean-100h~\cite{Panayotov2015} dataset to convert continuous frames to discrete codes. We quantize both learned representations with $K=100$ centroids. Leading to a bitrate of 700bps for CPC and 350bps for HuBERT.

% VQ-VAE
Similarly to CPC models, we trained the VQ-VAE content encoder model on the ``clean'' 6K hours subset from the LibriLight dataset. We use an encoder operating on the raw signal to extract discrete units, similar to~\cite{jukebox}. In addition, ``random restarts'' were performed when the mean usage of a codebook vector fell below a predetermined threshold. Finally, we used HiFiGAN (architecture and objective) as the decoder instead of a simple convolutional decoder, as it improved the overall audio quality. This model encodes the raw audio into a sequence of discrete tokens from 256 possible tokens~\cite{garbacea2019low} with a hop size of 160 raw audio samples. The VQ-VAE discrete code operates at a bitrate of 800bps. We additionally experimented with 100 discrete units for VQ-VAE, however results were the best for 256. This finding is consistent with~\cite{garbacea2019low}.

% verification model
The speaker verification network uses the architecture proposed in~\cite{heigold2016end}. It was trained on the VoxCeleb2~\cite{voxceleb2} dataset, achieving a 7.4\% Equal Error Rate (EER) for speaker verification on the test split of the VoxCeleb1~\cite{Nagrani17} dataset.

% pitch
Only a single F0 representation is considered across all evaluated models, trained on the VCTK dataset.
% The F0 is extracted from the raw audio using YAAPT~\cite{yaapt} algorithm, using a window size of 20ms and a 5ms hop. 
The F0 is extracted from the raw audio using a window size of 20ms and a 5ms hop. 
As a result, the F0 sequence is sampled at 200Hz. 
% We apply the quantization described at Sec.~\ref{sec:method}, using a pitch codebook of $K'=20$ tokens and an encoder that downsamples the pitch by $\times16$. 
The quantization described at Sec.~\ref{sec:method}, is applied using an F0 codebook of $K'=20$ tokens and an encoder that downsamples the signal by $\times16$. Hence, the discrete F0 representation is sampled at 12.5Hz, leading to a bitrate of 65bps. The final bitrate of the evaluated codecs is the sum of the pitch code bitrate with the content code bitrate.

% \paragraph*{Evaluation Metrics}
% \smallskip
\noindent{\bf Evaluation Metrics\quad} 
We consider both subjective and objective evaluation metrics. For subjective tests, we report the Mean Opinion Scores (MOS). In which human evaluators rate the naturalness of audio samples on a scale of 1--5. Each experiment, included 50 randomly selected samples rated by 30 raters. For objective evaluation, we consider: (i) Equal Error Rate~(EER) as an automatic speaker verification metric obtained using a pre-trained speaker verification network. We report EER between test utterances and enrolled speakers; (ii) Voicing Decision Error (VDE)~\cite{nakatani2008method}, which measures the portion of frames with voicing decision error; (iii) F0 Frame Error (FFE)~\cite{chu2009reducing}, measures the percentage of frames that contain a deviation of more than 20\% in pitch value or have a voicing decision error; (iv) Word Error Rate (WER) and Phoneme Error Rate (PER), proxy metrics to the intelligibility of the generated audio. We used a pre-trained ASR network~\cite{baevski2020wav2vec} on both reconstructed and converted samples to calculate both metrics. %To generate target phonemes, the g2p-en~\cite{g2pE2019} Grapheme2Phoneme module was used.

% \vspace{-0.1cm}
% \smallskip
\noindent{\bf Reconstruction \& Conversion}
% \vspace{-0.1cm}
We start by reporting the reconstruction performance. Results are summarized in Table~\ref{tab:recon}. When considering the intelligibility of the reconstructed signal HuBERT reaches the lowest PER and WER scores across all models, where both CPC and HuBERT are superior to VQ-VAE. However, when considering F0 reconstruction VQ-VAE outperforms both HuBERT and CPC by a significant margin. This results are somewhat intuitive, bearing in mind VQ-VAE objective is to fully reconstruct the input signal. In terms of subjective evaluation, all models reach similar MOS scores, with one exception of CPC on LJ. 

%Notice, since the same F0 units are used for each method, this result implies the VQ-VAE units contain some information about the F0 of the signal, enabling better reconstruction. Regarding speaker information, the CPC gets the lowest EER. 

To better evaluate the disentanglement properties of each method with respect to speaker identity and F0, we conducted an additional set of experiments aiming at speaker conversion and F0 manipulation. For voice conversion, we converted each test utterance into five random target speakers. Next, we employed a speaker verification network, which extracts \emph{d-vector} representation to evaluate speaker-converted utterances' similarity to real speaker utterances (low error-rate indicates good conversion), providing measurement to the speaker identity's disentanglement from the evaluated coding method. The error-rate is reported between converted test utterances and enrolled speakers. For the LJ speech single speaker dataset, we converted samples from the VCTK dataset to the single speaker and enrolled all VCTK speakers together with the single speaker. Results are summarized in Table~\ref{tab:conv} (left). Unlike resynthesis results, on voice conversion CPC and HuBERT outperform VQ-VAE on both LJ and VCTK datasets, indicating VQ-VAE contains more information about the speaker in the encoded units, hence producing more artifacts. Notice, this also affects WER, PER, and the overall subjective quality (MOS). 

Next, to evaluate the presence of F0 in the discrete units, we flattened the F0 units before synthesizing the signal and calculated VDE and FFE with respect to the original F0 values. F0 flattening was done by setting the speakers' mean F0 value across all voiced frames. In this experiment, we expected units that contain F0 information to be better at F0 reconstruction over disentangled units. Results are summarized in Table~\ref{tab:conv} (right). Notice VQ-VAE can still reconstruct the F0 almost at the same level as when using the original F0 as conditioning (5.2 vs 7.03, and 5.59 vs 7.8), in contrast to CPC and HuBERT.

\begin{figure}[t!]
\centering
\includegraphics[width=0.65\columnwidth, trim={50 20 70 20}]{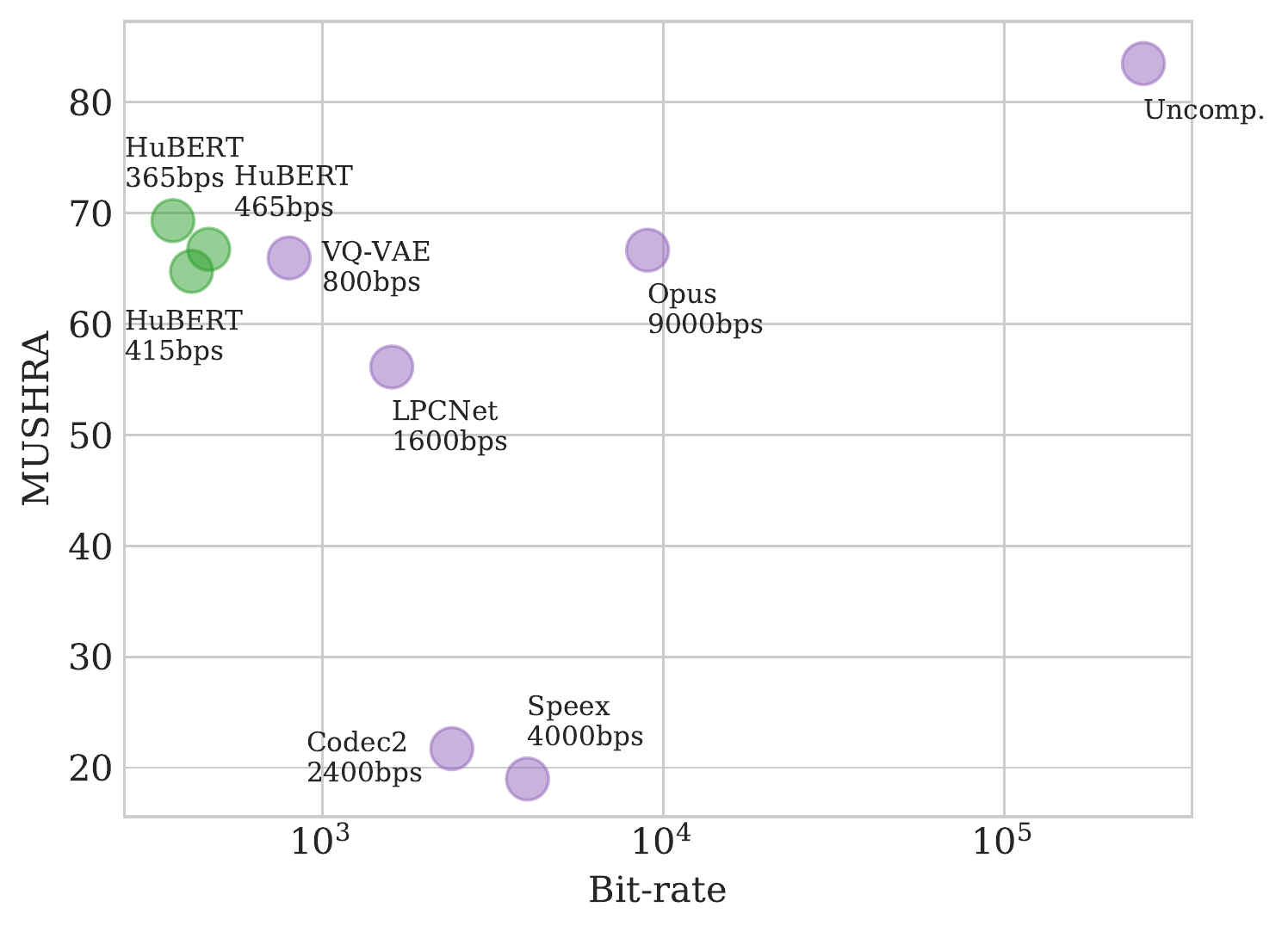}
% \caption{MUSHRA subjective listening test results as a function of bitrate per second for various methods. Purple dots denote the baseline methods, and green dots the proposed SSL based method.} 
\caption{MUSHRA subjective quality results as a function of bitrate per second. Purple dots denote the baseline methods, and green dots the proposed SSL based method.} 
\label{fig:codec}
\vspace{-0.5cm}
\end{figure}

% \vspace{-0.1cm}
% \smallskip
\noindent{\bf Speech Codec}
Our final experiment evaluates the obtained speech units as a low bitrate speech codec. 
% Therefore, we evaluate how the performance varies as a function of the number of discrete units. Changing the number of units is equivalent to varying the bitrate of the encoded signal. 
We use a subjective MUSHRA-type listening test~\cite{series2014method} to measure the perceived quality of the proposed speech codec with regard to its bitrate constraints. In MUSHRA evaluations, listeners are presented with a labeled uncompressed signal for reference, a set of test samples to rate, a copy of the uncompressed reference, and a low-quality anchor. Listeners are asked to rate each test utterance and the copy of the uncompressed reference with respect to the labeled reference in a scale of 1-100.

The experiment is performed on the VCTK dataset~\cite{vctk}. For evaluation, we used 20 utterances from 5 speakers. The set of speakers in the test data is disjoint with those in the training data. For this experiment, HuBERT models with 50, 100, and 200 units were trained as described in Sec.~\ref{sec:impl}. For comparison, we included other speech codecs in our evaluation: Opus~\cite{valin2012definition} wideband at 9 kbps VBR, Codec2~\cite{rowe2011codec} at 2.4 kbps and LPCNet~\cite{valin2019real} operating at 1.6 kbps. The LPCNet model was trained from scratch on the VCTK dataset following the experimental setup in~\cite{valin2019real}. The VQ-VAE model employs the HiFiGAN decoder trained on the LibriLight dataset to match the amount of data reported in~\cite{garbacea2019low}. We compressed the anchor sample with Speex~\cite{valin2016speex} at 4 kbps as a low anchor. Fig.~\ref{fig:codec} depicts the results. HuBERT with 50 units reaches the best MUSHRA score while its bitrate is only 365bps, which is significantly lower than the baseline methods.
\vspace{-0.3cm}
\section{Conclusion}
\vspace{-0.1cm}
\label{sec:conc}

We applied self-supervised discrete representations for the task of speech resynthesis. Furthermore, we demonstrated the efficiency of disentangled representations for signal reconstruction, voice conversion, and F0 manipulation. Our evaluations shed light on the properties encoded by each method in the context of speech synthesis. Finally, we adapt the HuBERT speech representation as an ultra lightweight speech codec, providing superior subjective results than the baselines with lower bitrate.

% For future work, we would like to extend the following method for expressive speech synthesis, including emotion and style conditioning. Moreover, we would like to improve the study on self-supervised speech codecs to be applied over in-the-wild and noisy-reverberent datasets. 

\bibliographystyle{IEEEtran}
\bibliography{bib}

\begin{thebibliography}{10}
\providecommand{\url}[1]{#1}
\csname url@samestyle\endcsname
\providecommand{\newblock}{\relax}
\providecommand{\bibinfo}[2]{#2}
\providecommand{\BIBentrySTDinterwordspacing}{\spaceskip=0pt\relax}
\providecommand{\BIBentryALTinterwordstretchfactor}{4}
\providecommand{\BIBentryALTinterwordspacing}{\spaceskip=\fontdimen2\font plus
\BIBentryALTinterwordstretchfactor\fontdimen3\font minus
  \fontdimen4\font\relax}
\providecommand{\BIBforeignlanguage}[2]{{%
\expandafter\ifx\csname l@#1\endcsname\relax
\typeout{** WARNING: IEEEtran.bst: No hyphenation pattern has been}%
\typeout{** loaded for the language `#1'. Using the pattern for}%
\typeout{** the default language instead.}%
\else
\language=\csname l@#1\endcsname
\fi
#2}}
\providecommand{\BIBdecl}{\relax}
\BIBdecl

\bibitem{oord2018representation}
A.~van~den Oord, Y.~Li, and O.~Vinyals, ``Representation learning with
  contrastive predictive coding,'' \emph{arXiv preprint arXiv:1807.03748},
  2018.

\bibitem{schneider2019wav2vec}
S.~Schneider, A.~Baevski, R.~Collobert, and M.~Auli, ``{wav2vec: Unsupervised
  Pre-Training for Speech Recognition},'' in \emph{INTERSPEECH}, 2019.

\bibitem{baevski2020wav2vec}
A.~Baevski \emph{et~al.}, ``wav2vec 2.0: A framework for self-supervised
  learning of speech representations,'' in \emph{ICLR}, 2020.

\bibitem{hsu2020hubert}
W.-N. Hsu \emph{et~al.}, ``Hubert: How much can a bad teacher benefit {ASR}
  pre-training?'' in \emph{NeurIPS Workshop on Self-Supervised Learning for
  Speech and Audio Processing Workshop}, 2020.

\bibitem{lakhotia2021generative}
K.~Lakhotia \emph{et~al.}, ``Generative spoken language modeling from raw
  audio,'' \emph{arXiv preprint arXiv:2102.01192}, 2021.

\bibitem{griffin1984signal}
D.~Griffin and J.~Lim, ``Signal estimation from modified short-time fourier
  transform,'' \emph{IEEE Transactions on Acoustics, Speech, and Signal
  Processing}, vol.~32, no.~2, pp. 236--243, 1984.

\bibitem{oord2016wavenet}
A.~v.~d. Oord \emph{et~al.}, ``Wavenet: A generative model for raw audio,''
  \emph{arXiv preprint arXiv:1609.03499}, 2016.

\bibitem{waveglow}
R.~Prenger, R.~Valle, and B.~Catanzaro, ``{Waveglow}: A flow-based generative
  network for speech synthesis,'' \emph{ICASSP}, 2019.

\bibitem{kong2020hifi}
J.~Kong \emph{et~al.}, ``Hifi-gan: Generative adversarial networks for
  efficient and high fidelity speech synthesis,'' in \emph{NeurIPS}, 2020.

\bibitem{ebbers2017hidden}
J.~Ebbers \emph{et~al.}, ``Hidden markov model variational autoencoder for
  acoustic unit discovery,'' in \emph{INTERSPEECH 2017}, 2017.

\bibitem{van2017neural}
A.~van~den Oord, O.~Vinyals \emph{et~al.}, ``Neural discrete representation
  learning,'' in \emph{NeurIPS}, 2017.

\bibitem{hsu2017unsupervised}
W.-N. Hsu, Y.~Zhang, and J.~Glass, ``Unsupervised learning of disentangled and
  interpretable representations from sequential data,'' in \emph{Advances in
  Neural Information Processing Systems}, 2017.

\bibitem{kreuk2020self}
F.~Kreuk, J.~Keshet, and Y.~Adi, ``Self-supervised contrastive learning for
  unsupervised phoneme segmentation,'' \emph{arXiv preprint arXiv:2007.13465},
  2020.

\bibitem{devlin-etal-2019-bert}
J.~Devlin \emph{et~al.}, ``{BERT}: Pre-training of deep bidirectional
  transformers for language understanding,'' in \emph{NAACL}, 2019.

\bibitem{prenger2019waveglow}
R.~{Prenger}, R.~{Valle}, and B.~{Catanzaro}, ``Waveglow: A flow-based
  generative network for speech synthesis,'' in \emph{ICASSP}, 2019.

\bibitem{dunbar2019zero}
E.~Dunbar \emph{et~al.}, ``{The Zero Resource Speech Challenge 2019: TTS
  Without T},'' in \emph{Proc. Interspeech 2019}, 2019, pp. 1088--1092.

\bibitem{dunbar2020zero}
------, ``{The Zero Resource Speech Challenge 2020: Discovering Discrete
  Subword and Word Units},'' in \emph{Proc. Interspeech 2020}, 2020, pp.
  4831--4835.

\bibitem{eloff2019unsupervised}
R.~Eloff \emph{et~al.}, ``{Unsupervised Acoustic Unit Discovery for Speech
  Synthesis Using Discrete Latent-Variable Neural Networks},'' in
  \emph{INTERSPEECH}, 2019.

\bibitem{jin2018fftnet}
Z.~{Jin}, A.~{Finkelstein}, G.~J. {Mysore}, and J.~{Lu}, ``Fftnet: A real-time
  speaker-dependent neural vocoder,'' in \emph{ICASSP}, 2018.

\bibitem{tjandra2020transformer}
A.~Tjandra, S.~Sakti, and S.~Nakamura, ``{Transformer VQ-VAE for Unsupervised
  Unit Discovery and Speech Synthesis: ZeroSpeech 2020 Challenge},'' in
  \emph{INTERSPEECH}, 2020.

\bibitem{trans}
A.~Vaswani \emph{et~al.}, ``Attention is all you need,'' in \emph{NeurIPS},
  2017.

\bibitem{van2020vector}
B.~van Niekerk, L.~Nortje, and H.~Kamper, ``{Vector-Quantized Neural Networks
  for Acoustic Unit Discovery in the ZeroSpeech 2020 Challenge},'' in
  \emph{INTERSPEECH}, 2020.

\bibitem{zhao2020improved}
Y.~Zhao \emph{et~al.}, ``Improved prosody from learned f0 codebook
  representations for vq-vae speech waveform reconstruction,'' \emph{arXiv
  preprint arXiv:2005.07884}, 2020.

\bibitem{polyak2019tts}
A.~Polyak, L.~Wolf, and Y.~Taigman, ``{TTS Skins: Speaker Conversion via
  ASR},'' in \emph{INTERSPEECH}, 2020.

\bibitem{polyak2020unsupervised}
A.~Polyak \emph{et~al.}, ``{Unsupervised Cross-Domain Singing Voice
  Conversion},'' in \emph{INTERSPEECH}, 2020.

\bibitem{polyak2021high}
------, ``High fidelity speech regeneration with application to speech
  enhancement,'' \emph{ICASSP}, 2021.

\bibitem{atal1971speech}
B.~S. Atal and S.~L. Hanauer, ``Speech analysis and synthesis by linear
  prediction of the speech wave,'' \emph{The journal of the acoustical society
  of America}, vol.~50, no.~2B, pp. 637--655, 1971.

\bibitem{griffin1985new}
D.~Griffin and J.~Lim, ``A new model-based speech analysis/synthesis system,''
  in \emph{ICASSP}, 1985.

\bibitem{mccree19962}
A.~McCree, K.~Truong, E.~B. George, T.~P. Barnwell, and V.~Viswanathan, ``A 2.4
  kbit/s melp coder candidate for the new us federal standard,'' in
  \emph{ICASSP}, 1996.

\bibitem{kleijn2018wavenet}
W.~B. Kleijn, F.~S. Lim, A.~Luebs, J.~Skoglund, F.~Stimberg, Q.~Wang, and T.~C.
  Walters, ``Wavenet based low rate speech coding,'' in \emph{ICASSP}, 2018.

\bibitem{lim2020robust}
F.~S. Lim \emph{et~al.}, ``Robust low rate speech coding based on cloned
  networks and wavenet,'' in \emph{ICASSP}, 2020.

\bibitem{valin2019real}
J.-M. Valin and J.~Skoglund, ``A real-time wideband neural vocoder at 1.6 kb/s
  using lpcnet,'' \emph{arXiv preprint arXiv:1903.12087}, 2019.

\bibitem{valin2019lpcnet}
------, ``Lpcnet: Improving neural speech synthesis through linear
  prediction,'' in \emph{ICASSP}, 2019.

\bibitem{garbacea2019low}
C.~G{\^a}rbacea \emph{et~al.}, ``Low bit-rate speech coding with vq-vae and a
  wavenet decoder,'' in \emph{ICASSP}, 2019.

\bibitem{skoglund2019improving}
J.~Skoglund and J.-M. Valin, ``Improving opus low bit rate quality with neural
  speech synthesis,'' \emph{arXiv preprint arXiv:1905.04628}, 2019.

\bibitem{valin2012definition}
J.-M. Valin, K.~Vos, and T.~Terriberry, ``Definition of the opus audio codec,''
  \emph{IETF, September}, 2012.

\bibitem{kingma2013auto}
D.~P. Kingma and M.~Welling, ``Auto-encoding variational bayes,'' \emph{ICLR},
  2014.

\bibitem{yaapt}
K.~{Kasi} and S.~A. {Zahorian}, ``Yet another algorithm for pitch tracking,''
  \emph{ICASSP}, 2002.

\bibitem{jukebox}
P.~Dhariwal, H.~Jun, C.~Payne, J.~W. Kim, A.~Radford, and I.~Sutskever,
  ``Jukebox: A generative model for music,'' \emph{arXiv preprint
  arXiv:2005.00341}, 2020.

\bibitem{heigold2016end}
G.~Heigold, I.~Moreno, S.~Bengio, and N.~Shazeer, ``End-to-end text-dependent
  speaker verification,'' in \emph{ICASSP}, 2016.

\bibitem{larsen2016autoencoding}
A.~B.~L. Larsen \emph{et~al.}, ``Autoencoding beyond pixels using a learned
  similarity metric,'' in \emph{ICML}, 2016.

\bibitem{ljspeech17}
K.~Ito and L.~Johnson, ``The lj speech dataset,''
  \url{https://keithito.com/LJ-Speech-Dataset/}, 2017.

\bibitem{vctk}
C.~Veaux \emph{et~al.}, ``{CSTR} {VCTK} {Corpus}: English multi-speaker corpus
  for {CSTR} voice cloning toolkit,'' 2017.

\bibitem{Riviere2020}
M.~Rivi{\`e}re and E.~Dupoux, ``Towards unsupervised learning of speech
  features in the wild,'' in \emph{SLT 2020: IEEE Spoken Language Technology
  Workshop}, 2020.

\bibitem{Kahn2020}
J.~{Kahn} \emph{et~al.}, ``Libri-light: A benchmark for asr with limited or no
  supervision,'' in \emph{ICASSP}, 2020.

\bibitem{Panayotov2015}
V.~Panayotov, G.~Chen, D.~Povey, and S.~Khudanpur, ``Librispeech: an asr corpus
  based on public domain audio books,'' in \emph{ICASSP}, 2015.

\bibitem{voxceleb2}
J.~S. Chung, A.~Nagrani, and A.~Zisserman, ``Voxceleb2: Deep speaker
  recognition,'' \emph{INTERSPEECH}, 2018.

\bibitem{Nagrani17}
A.~Nagrani, J.~S. Chung, and A.~Zisserman, ``Voxceleb: a large-scale speaker
  identification dataset,'' \emph{INTERSPEECH}, 2017.

\bibitem{nakatani2008method}
T.~Nakatani \emph{et~al.}, ``A method for fundamental frequency estimation and
  voicing decision: Application to infant utterances recorded in real
  acoustical environments,'' \emph{Speech Communication}, 2008.

\bibitem{chu2009reducing}
W.~Chu and A.~Alwan, ``Reducing f0 frame error of f0 tracking algorithms under
  noisy conditions with an unvoiced/voiced classification frontend,''
  \emph{ICASSP}, 2009.

\bibitem{series2014method}
B.~Series, ``Method for the subjective assessment of intermediate quality level
  of audio systems,'' \emph{International Telecommunication Union
  Radiocommunication Assembly}, 2014.

\bibitem{rowe2011codec}
D.~Rowe, ``Codec 2-open source speech coding at 2400 bits/s and below,'' in
  \emph{TAPR and ARRL 30th Digital Communications Conference}, 2011, pp.
  80--84.

\bibitem{valin2016speex}
J.-M. Valin, ``Speex: A free codec for free speech,'' \emph{arXiv preprint
  arXiv:1602.08668}, 2016.

\end{thebibliography}

\end{document}